\documentclass[useAMS, usenatbib]{mn2e}

\usepackage{amsmath}
\usepackage{graphicx}
\usepackage{natbib}

\include{xcolor}

\usepackage{hyperref}
\hypersetup{
    colorlinks=true,
    linkcolor=black,
    citecolor=black,
    filecolor=blue,
    urlcolor=blue
}

\newcommand{\zobov}{\textsc{ZOBOV}}

\begin{document}

\title[Voids in Ly$\alpha$ Forest Tomographic Maps]{
Finding High-Redshift Voids Using Lyman-$\alpha$ Forest Tomography
}
\author[Stark et al.]{
Casey W. Stark$^{1}$\thanks{email: caseywstark@berkeley.edu},
Andreu Font-Ribera$^{2}$,
Martin White$^{1, 2, 3}$,
Khee-Gan Lee$^{4}$ \\
$^1$ Department of Astronomy, University of California, Berkeley, CA 94720,
USA \\
$^2$ Lawrence Berkeley National Laboratory, 1 Cyclotron Road, Berkeley, CA
93720, USA \\
$^3$ Department of Physics, University of California, Berkeley, CA 94720, USA \\
$^4$ Max Planck Institute for Astronomy, K\"{o}nigstuhl 17, D-69117 Heidelberg,
West Germany
}
\date{\today}
\pagerange{\pageref{firstpage}--\pageref{lastpage}}

\maketitle

\label{firstpage}

\begin{abstract}
We present a new method of finding cosmic voids using tomographic maps of
Ly$\alpha$ forest flux.
We identify cosmological voids with radii of 2 -- $12 \, h^{-1}$Mpc in a large
N-body simulation at $z = 2.5$, and characterize the signal of the high-redshift
voids in density and Ly$\alpha$ forest flux.
The void properties are similar to what has been found at lower redshifts, but
they are smaller and have steeper radial density profiles.
Similarly to what has been found for low-redshift voids, the radial velocity
profiles have little scatter and agree very well with the linear theory
prediction.
We run the same void finder on an ideal Ly$\alpha$ flux field and
tomographic reconstructions at various spatial samplings.
We compare the tomographic map void catalogs to the density void catalog and
find good agreement even with modest-sized voids ($r > 6 \, h^{-1}$Mpc).
Using our simple void-finding method, the configuration of the ongoing
CLAMATO survey covering $1\,$deg$^2$ would provide a sample of about 100
high-redshift voids.
We also provide void-finding forecasts for larger area surveys, and discuss
how these void samples can be used to test modified gravity models, study
high-redshift void galaxies, and to make an Alcock-Paczynski measurement.
To aid future work in this area, we provide public access to our simulation
products, catalogs, and sample tomographic flux maps.
\end{abstract}


\section{Introduction}

The material we see in the Universe around us makes up a beaded,
filamentary network known as the ``cosmic web'' \citep{Bon96}
This web appears to be the natural outcome of gravitational instability
acting upon an initially Gaussian random field.
The majority of the cosmic web, by volume, is made up of large,
almost empty regions known as voids which are surrounded by walls,
filaments and clusters \citep[see e.g.][for a review]{Wey11a}.
Within this paradigm, voids are regions which are practically devoid of
galaxies.
They are slightly prolate in shape and occur on a wide range
of sizes from Mpc to tens of Mpc \citep{Vog94, Cec06, Lav12}.

The study of cosmic voids has received renewed theoretical attention recently.
Voids are intrinsically interesting as a major constituent of the Universe
(by volume) and one of the most visually striking features in galaxy maps.
They form an interesting environment for the study of galaxy evolution.
They may present an excellent laboratory for studying material which
clusters most weakly (e.g.\ dark energy or massive neutrinos),
and for testing modified gravity models.
Future surveys are expected to find large samples of voids at a range of
redshifts, enhancing the potential of void science.

The pristine environments of voids present an interesting setting for the study
of early galaxy formation.
Galaxies in low-redshift voids generally have smaller stellar masses,
appear bluer, have a later morphological type, and have higher specific star
formation rates than galaxies in average density environments
\citep{Wey11a, Bey15}, although the latter properties might be solely due to
their lower stellar mass \citep{Hoy05, Kre11}.
Extending similar studies to higher redshifts to see whether similar trends
hold is a pressing observational challenge.

\citet{Ryd95} was the first to discuss using voids as probes of cosmology.
\citet{Par07} anticipated using void ellipticity as a cosmological probe
and \citet{Lee09, Bos12} discussed constraining dark energy using voids.
\citet{Lav12} investigated the potential for using stacked voids as a probe
of geometrical distortions \cite[the AP test;][]{Alc79}.
\citet{Cha14} have studied the clustering of voids and \citet{Ham14c} describe
constraining cosmology with void-galaxy cross-correlations.
\citet{Hel10, Li11} have investigated studying the nature of dark matter using
the properties of voids and \citet{Li12, Cla13} have suggested that void
properties may provide a strong test of some modified gravity theories.

Observationally, studies of voids date back over three decades
\citep{Gre78, Lon78, Kir81}.
Recent redshift surveys have identified large samples of voids
(e.g.\ 2dF: \citealt{Cec06}; SDSS: \citealt{Sut12, Sut14a};
VIPERS: \citealt{Mic14}) and a measurement of the AP effect from voids in the
local Universe has recently been reported by \citet{Sut14b, Ham14a}.
Being underdense in both galaxies and dark matter, voids act like objects
with an effectively negative mass, bending light rays away from them.
This effect has been recently detected at high significance by \citet{Cla14}.

In the absence of large-scale dynamical and environmental influences,
voids would become increasingly isotropic as they evolve \citep{Ick84}.
However, in modern theories of structure formation the frequent encounters
with surrounding structures and the influence of large-scale tidal fields
serve to reverse the simple trend expected for isolated voids \citep{Wey11a}.
As matter in the center of voids streams outwards faster than matter
towards the boundary, the interior evolves into an almost uniform low density
region surrounded by `ridges' marking the void edge: often referred to as
a ``bucket-shaped'' density profile \citep[see][for recent fits]{Cec06, Ham14b}.
The density in the center has a characteristic value of $\delta \approx -0.8$.

Historically, surveys of voids over large volumes have come from large,
galaxy redshift surveys.
However, finding voids in this manner requires a significant investment in
telescope time due to the necessity of a high spatial sampling of tracer
galaxies.
For example, the void catalog presented in \citet{Sut12} found voids in the
distribution of SDSS DR7 galaxies.
Their `bright' cut found voids with radii larger than $7 \, h^{-1}$Mpc with
galaxies separated by $8 \, h^{-1}$Mpc.
To find comparable galaxy separations at $z = 0.5$, 1.0, and 2.0 will require
obtaining complete galaxy redshift samples for apparent limiting magnitudes of
$I = 22.5$, 24.2, and 25.7, respectively (assuming galaxy luminosity functions
from \citealt{Dah05} at $z \leq 1$, and \citealt{Red08} at $z = 2$).
So while such galaxy number densities are just achievable up to $z \approx 1$
with existing telescopes, it becomes increasingly challenging at higher
redshifts.

In light of the aforementioned challenges, it is understandable that little
attention has been given to studying voids at $z > 1$
\citep[although see][]{Dal07, Vie08}.
However, recently it has been noted that given sufficient sightlines, the
Ly$\alpha$ forest observed in a dense grid of faint background galaxies and
quasars can be used to create three-dimensional maps of large-scale structure
and that the observational requirements to map out cosmological volumes
($V \approx 10^6 \, h^{-3} \mathrm{Mpc}^3$) are within reach of existing
facilities \citep{Lee14a}.
Indeed, the first pilot map on a small field has already been made using
just a few hours of data from Keck telescope \citep{Lee14b}.
This method is ideally suited to finding extended structures at high
redshift.
In \citet{Sta14}, we investigated the possibility of finding
protoclusters in tomographic Ly$\alpha$ maps.
Here, we study the signature of cosmological voids in the Ly$\alpha$ forest.

The outline of the paper is as follows.
In \S\ref{sec:sim}, we briefly describe the simulations that we use in
this paper and the method we use to find voids.
In \S\ref{sec:voids}, we use our catalog of high-redshift voids to explore
the properties of voids in density and flux.
In \S\ref{sec:flux}, we demonstrate how to find voids using tomographic flux
maps and discuss how well voids found in the maps compare to those found in
the matter density.
In \S\ref{sec:discussion}, we present the prospects of finding high-redshift
voids with this method in ongoing and future surveys, and discuss cosmology
applications.
We present our conclusions in \S\ref{sec:conclusions}.

\section{Simulations and void finding}
\label{sec:sim}

\subsection{N-body simulations}

In order to study the signal of voids in the Ly$\alpha$
forest, we make use of cosmological $N$-body simulations.
We require a simulation which simultaneously covers a large cosmological
volume while having a sufficiently small inter-particle spacing to model
transmission in the IGM. The requirements are sufficiently demanding that
we have used a pure $N$-body simulation, augmented with the fluctuating
Gunn-Peterson approximation \citep[FGPA;][]{Pet95, Cro98, Mei01, Mei09}.
This same simulation was also used in \citet{Lee14a} and \citet{Sta14} so we
only review the salient features here.
We are also providing public access to the relevant simulation products --
see App.~\ref{app:data} for more details.

The simulation employed $2560^3$ equal mass
($8.6 \times 10^7 \, h^{-1} M_\odot$) particles in a $256 \, h^{-1}$Mpc
periodic, cubical box.
This provides sufficient mass resolution to model the large-scale features in
the IGM at $z = 2$--3 using the FGPA
\citep{Mei01, Ror13} and sufficient volume to find
large voids.
The assumed cosmology was of the flat $\Lambda$CDM family,
with $\Omega_{\rm m} \approx 0.31$, $\Omega_{\rm b} h^2 \approx 0.022$,
$h = 0.6777$, $n_s = 0.9611$, and $\sigma_8 = 0.83$, in agreement with
\citet{planck_2013_XVI}.
The initial conditions were generated using second-order Lagrangian
perturbation theory at $z_{\rm ic} = 150$, when the rms particle
displacement was 40 per cent of the mean inter-particle spacing.
The particle positions and velocities were evolved using the TreePM code of
\citet{Whi02}.
Throughout the text, we will use the particle positions and
velocities from the output at $z = 2.5$. Using the particle positions and
velocities at $z = 2.5$, we generated mock Ly$\alpha$ forest
spectra on a $2560^3$ grid with the FGPA as described in
\citet{Sta14}. In all, we generated $2560^3$ grids with the matter density and
Ly$\alpha$ forest flux in real- and redshift-space.
For many purposes in this work, we did not need the high resolution provided by
the $2560^3$ grids and found it much easier to work with smaller grids.
For this reason, we also downsampled the fields to $256^3$ by simply averaging
neighboring grid points.
In the remainder of the paper, when we refer to flux,
we mean the Ly$\alpha$ forest transmitted flux fraction perturbation

\begin{equation}
  \delta_F = F / \langle F \rangle - 1 ~.
\end{equation}

\subsection{Void finding}
\label{subsec:void_finding}

\begin{figure*}
\begin{center}
\includegraphics[width=\textwidth]{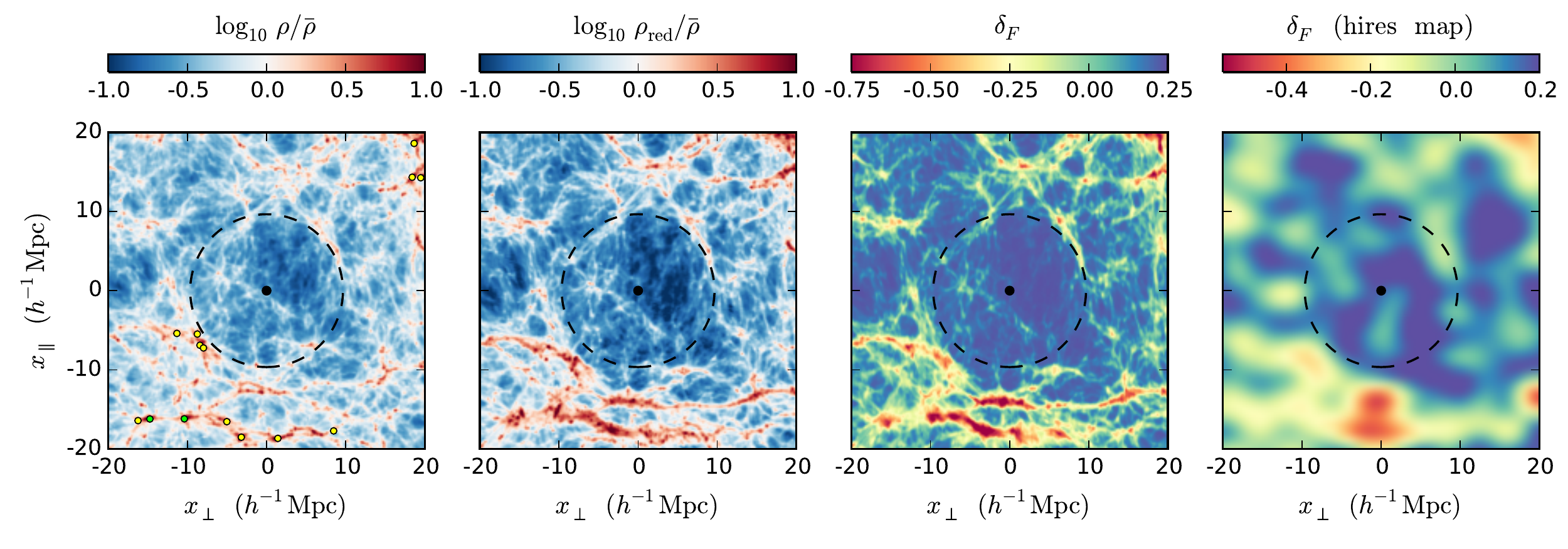}
\end{center}
\caption{
A slice through our simulation and a mock reconstructed flux map centered on a
large void, with a radius of $9.7 \, h^{-1}$Mpc.
The slice is $40 \, h^{-1}$Mpc across and $6 \, h^{-1}$Mpc into the page.
The vertical axis, $x_\parallel$, is the redshift direction,
along the line of sight, and the horizontal axis, $x_\perp$,
is one of the directions transverse to the line of sight (the other transverse
direction is into the page).
(Left) The matter density, in real space.
The void center and radius are indicated with a black dot and dashed circle.
Positions of halos with $M \geq 10^{12} \, h^{-1} M_\odot$ are plotted as green
dots and positions of halos with $M \geq 3 \times 10^{11} \, h^{-1} M_\odot$ are
plotted as yellow dots.
These halos would host galaxies at the spectroscopic limits for existing
instrumentation.
(Middle left) The matter density, in redshift space.
(Middle right) The `true' flux field over the same volume, in redshift space.
(Right) A reconstructed flux field from a mock survey (also in redshift space)
with average sightline spacing, $d_\perp=2.5 \, h^{-1}$Mpc (see text).
}
\label{fig:slice}
\end{figure*}

There are a variety of methods and tools used to find voids in large-scale
structure \citep{Kau91, Pla07, Ney07, Sut15}.
We use a simple spherical underdensity method on the low resolution, gridded
densities to construct our $z = 2.5$ void catalog.
This technique is similar to spherical overdensity (SO) halo finding, but
instead applied to underdensities.
We identify voids by taking the $256^3$ density grid and selecting points under
some threshold value, then growing spheres around the points until the
average density enclosed reaches a target value.
We handle overlapping voids by only saving the void with the largest radius,
and we also discard any remaining voids with a radius $r < 2 \, h^{-1}$Mpc.

The free parameters in this void-finding method are the threshold value
$\rho_{\rm thresh}$ and enclosed average target value $\bar{\rho}_{\rm enc}$.
For the density field, there is a well-motivated threshold value of
$\rho_{\rm thresh} = 0.2 \, \bar \rho$.
This threshold value is a canonical density for a void core, corresponding to
the central density at shell-crossing for spherical void models \citep{Wey11a}.
This critical density value is a standard choice in other void-finding codes
\citep[e.g.][]{Ney07}.
The choice of the average target value, however, is somewhat arbitrary.
We first tested an average target value of
$\bar{\rho}_{\rm enc} = 0.2 \, \bar \rho$, but found that it produced voids that
were far too small -- the spheres never reached the apparent `edge' surrounding
the low-density core.
We experimented with several other average target values and found that a value
of $\bar{\rho}_{\rm enc} =0.4 \, \bar \rho$ resulted in good agreement between
the sphere sizes and the apparent void edges.

In principle, there is nothing special about the specific threshold and average
density values we chose, and these parameters should depend on the redshift.
That is, as voids continue to evacuate, the core and average densities of the
void will decrease.
In practice, we found the final void catalog is not very sensitive to these
settings, although the void radii clearly scale with the average target density
setting.
Since most large voids have central densities $< 0.2 \, \bar \rho$ already, the
exact value of the threshold mostly makes a difference in terms of how many
points we must search over, and less of a difference in the void centers.
We did find that a very small threshold (say $< 0.1 \, \bar \rho$ for this
redshift) forces voids to grow from positions that often look off-center by eye.
This is due to the hierarchical nature of voids, in that the lowest-density
point in a large void is typically the center of a smaller subvoid, sometimes
referred to as the void-in-void scenario (c.f.\ fig.~6 of \citealt{Ney07} and
\citealt{She04}).
Using the SO parameter values of the threshold
$\rho_{\rm thresh} = 0.2 \, \bar \rho$ and average target
$\bar{\rho}_{\rm enc} = 0.4 \, \bar \rho$, we found 16,167 voids which cover
15 per cent of the simulation volume.

\begin{figure}
\begin{center}
\includegraphics[width=\columnwidth]{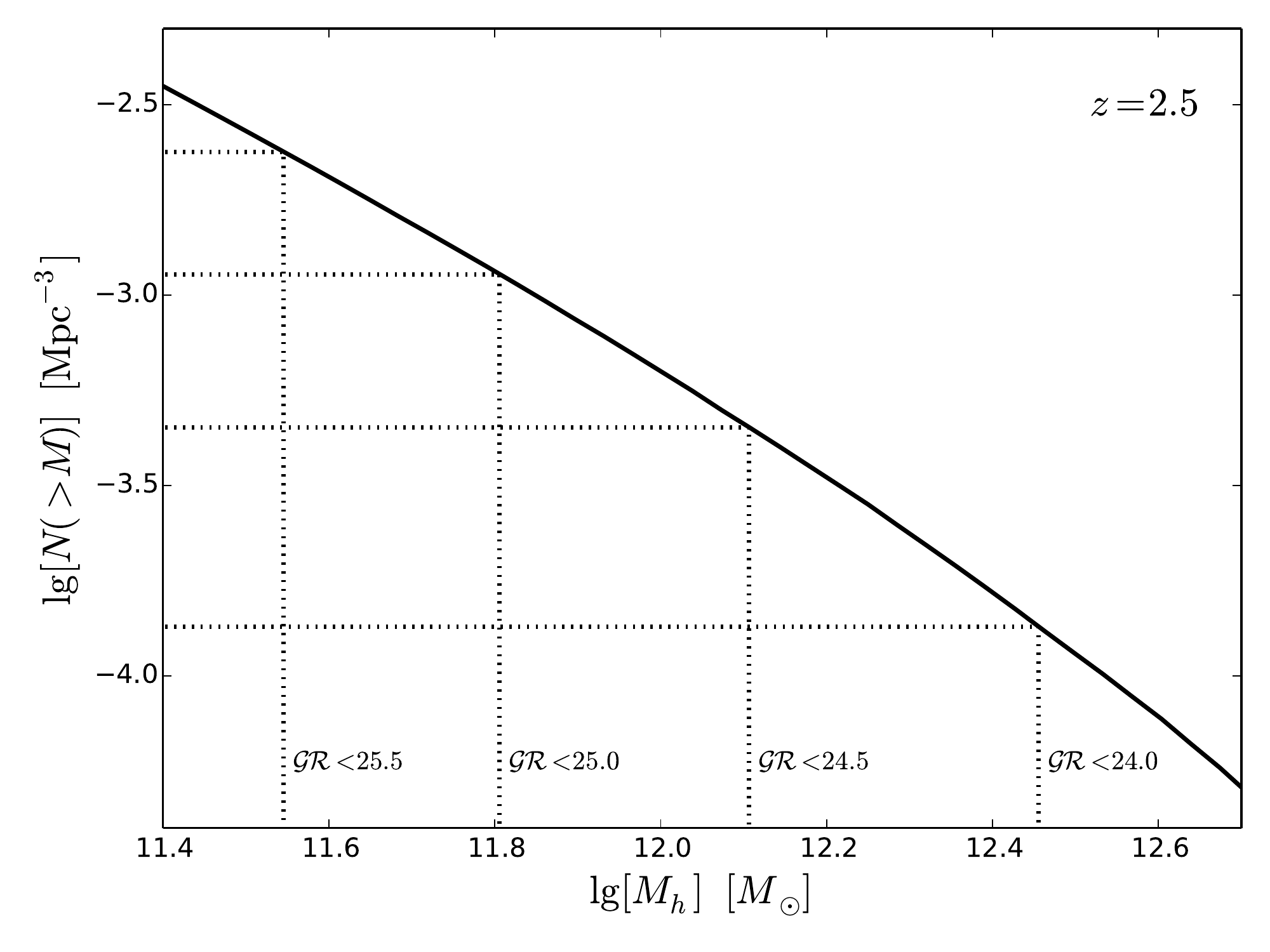}
\end{center}
\caption{
The cumulative, halo mass function in the simulation at $z = 2.5$,
as a function of (FoF) halo mass.
The dotted lines indicate the abundances of galaxies brighter than the listed
apparent $\mathcal{R}$-magnitude limits, derived from the luminosity function
of \protect\citet{Red08}.
Since at this redshift and these masses satellites make up a small fraction of
galaxies by number, this plot allows us to approximately equate our mass-limited
halo catalogs into flux-limited galaxy catalogs.
Note we have used volumes and masses without factors of $h$ in this figure to
match the scalings adopted in \protect\citet{Red08}.
}
\label{fig:massfn}
\end{figure}

Fig.~\ref{fig:slice} shows a slice through our simulation, centered on one
of the largest voids in our catalog with $r = 9.7 \, h^{-1}$Mpc.
The slice is $40 \, h^{-1}$Mpc across and $6 \, h^{-1}$Mpc projected into the
page.
The four panels show the void in real-space density, redshift-space density,
Ly$\alpha$ forest flux, and a tomographic flux map constructed from a mock
survey.
In each image, we marked the void center and radius with a black dot and dashed
line, respectively.
In the first panel, we also overplotted the positions of halos found in the
same slice.
Green dots mark the positions of halos with mass
$M \geq 10^{12} \, h^{-1} M_\odot$ (roughly an $L_\star$ halo at this redshift),
while yellow dots mark the positions of halos
with $3 \times 10^{11} \, h^{-1} M_\odot \le M < 10^{12} \, h^{-1} M_\odot$.
Based on simple abundance matching (see Fig.~\ref{fig:massfn}),
these halos should host galaxies with apparent magnitudes $\mathcal{R} < 24.7$
and $24.7 \leq \mathcal{R} < 25.6$, just bright enough for redshift
determination with existing facilities.
The second panel shows that in redshift space the void has a larger density
contrast and extent in the line-of-sight direction due to the outflow of
matter from the void.
Such a large structure is easily visible in the redshift-space density and flux.
Although the tomographic flux map is a blurred version of the true flux,
the void structure is so large that it can still easily be picked out by eye.
For reference, the tomographic map is one of the realizations from
\citet{Sta14} with an average sightline spacing of
$d_\perp = 2.5\,h^{-1}\mathrm{Mpc}$, similar to the ongoing survey of
\citet{Lee14b}.
At the same time, the void is not captured by the galaxy positions even if we
assume a complete galaxy sample.
The relative sparsity of such halos highlights the difficulty in finding
voids, even large ones, at high redshift using galaxies as tracers.

We compared our void catalog with that produced by a watershed void finder,
similar to the \zobov\ code.
The watershed method finds the set of connected elements all under some
threshold.
In \zobov\, the elements are the Voronoi cells in the tessellation of the dark
matter particle positions (where the density is estimated from the volume of the
Voronoi cell), but in this case, we use the density values on the $2560^3$ grid
for simplicity.
The watershed algorithm on a uniform grid is straightforward.
We find the set of points under the given threshold, and keep a list of
the under-threshold points that have not been assigned to a specific watershed.
Starting with the minimum value point, we search grid neighbors to see if they
are also under the threshold and add them to the current watershed if so.
The search stops when there are no remaining neighbors under the threshold.
These points are then removed from the list of unassigned points and we move on
to the next watershed.
After we assign all points under the threshold, we discard watersheds with an
effective radius $r_{\rm eff} = (3V_{\rm shed}/4\pi)^{1/3}$ less than
$2 \, h^{-1}$Mpc, as we did with the spherical underdensity voids.
Using this method with the same threshold of $\rho < 0.2 \, \bar \rho$,
we found 6,364 voids, covering 5 per cent of the simulation volume.

The sets of large voids in the spherical underdensity (or SO) catalog and
the watershed catalog agree very well.
We visually inspected the 100 largest voids in the SO catalog, and found
matches in the watershed catalog.
In most cases the watershed void effective radius was slightly smaller
(by 1--$2\,h^{-1}$Mpc), which explains the total count and volume difference,
and the watershed voids typically have complex morphologies.
The watershed voids often have an ellipsoidal core, with fingers stretching out
into smaller low-density regions.
We compared the SO void centers to the watershed void value-weighted
centroids
$\vec{x}_{\rm shed} = \sum_i \vec{x}_i \rho_i^{-1} / \sum_i \rho_i^{-1}$,
where the sums are over all the points in the shed, and we weight by the inverse
of the density so that the centering is driven by lower-density points.
Unfortunately, the non-spherical geometries of the watershed voids tend to drive
the centroid away from the center found with the SO method and the centers in
the two catalogs tend to disagree by several Mpc
(see Appendix~\ref{app:sheds} for more discussion and images of the watershed
voids).
It is reassuring that these two methods for finding voids in the density
field qualitatively agree well, but we decided to use the SO void catalog for
the remainder of this work due to its simplicity.
Overall, the centers and simple shapes of the SO voids provide cleaner radial
profiles and should be easier to find in the tomographic maps later.
We were also concerned that the non-trivial noise we expect in the maps from
tomographic reconstruction might artificially combine or split watershed
regions, whereas the spherical average in the SO method will be less affected
by such noise.
Since our tomographic maps come with a noise estimate, one could imagine a more
sophisticated algorithm (e.g.\ a matched filter or likelihood-based method) for
finding voids could be implemented.
We leave such investigations to future work.

\section{Voids at \lowercase{$z = 2.5$}}
\label{sec:voids}

\begin{figure}
\begin{center}
\includegraphics[width=\columnwidth]{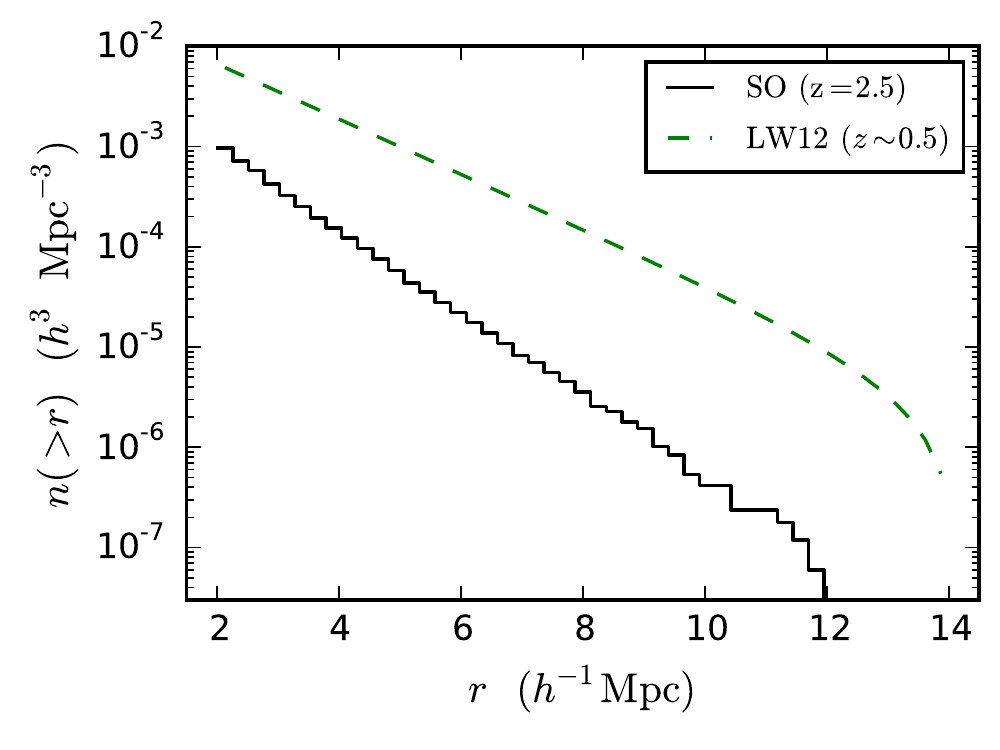}
\end{center}
\caption{
The cumulative, comoving number density of voids with radii greater than
$r$ vs.\ $r$.
In black, we show the distribution of our SO density voids at $z = 2.5$.
The green dashed line is the distribution of the voids found in
\protect\citet{Lav12}, which are generally larger, as expected.
}
\label{fig:radius_dist}
\end{figure}

In Fig.~\ref{fig:radius_dist}, we show the cumulative number density of voids
as a function of void radius.
We plot the distribution of our $z = 2.5$ voids in black, and show the
distribution of low-redshift voids with a green dashed line
($z \approx 0.5$), computed from eq.~21 of \citet{Lav12}.
As expected, there are many more small voids and voids are generally smaller at
$z = 2.5$ than at $z \simeq 0$
(c.f.\ fig.~1 of \citealt{Cec06} or fig.~7 of \citealt{Lav12}).
While voids with radii of $7 \, h^{-1}$Mpc are common for low-redshift
studies, we have only 126 voids with $r \ge 7 \, h^{-1}$Mpc, which cover
two per cent of the simulation volume.
We note, however, that it is difficult to compare void sizes across works using
different void-finding methods and working at different redshifts.
For instance, we could increase the number of $r \ge 7 \, h^{-1}$Mpc voids by
simply increasing the average target value in our SO void finder.
For the most part, this does not change which large voids are identified,
but does shift centers and increase the cumulative number density at a
particular value.
See \citet{Col08} for more detail about the difficulties of defining voids and
differing results from various void finders.

\begin{figure}
\begin{center}
\includegraphics[width=\columnwidth]{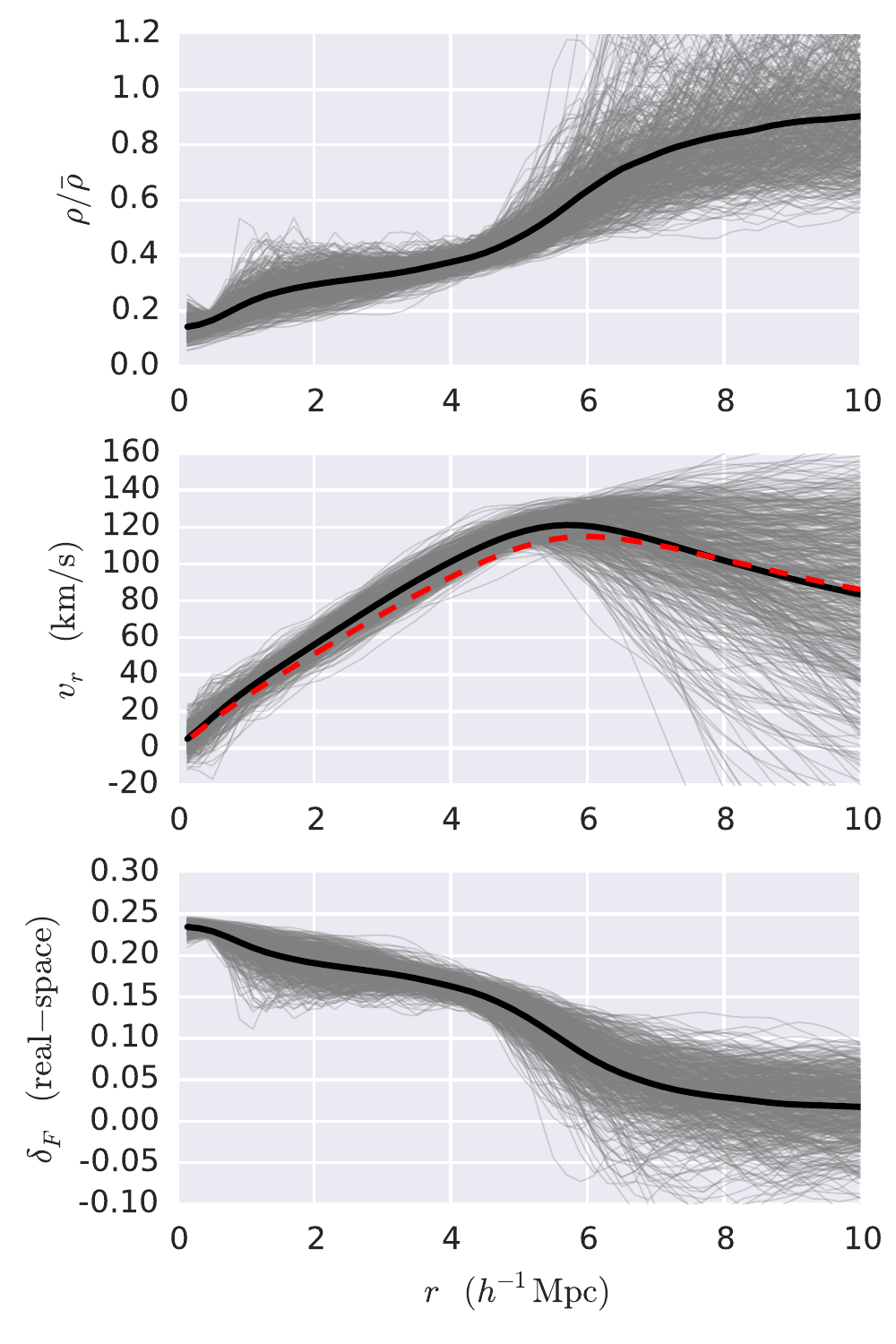}
\end{center}
\caption{
Radial profiles centered on voids with $5 \leq r < 6 \, h^{-1}$Mpc.
Gray lines are individual profiles, and the thick black line shows the mean.
(Top) The matter density profile.
(Middle) The radial peculiar velocity profile. The dashed red line shows the
linear theory prediction using the density stack profile above.
(Bottom) The real-space flux profile.
}
\label{fig:profile}
\end{figure}

Fig.~\ref{fig:profile} shows the radial profile of voids in density, radial
peculiar velocity, and real-space flux, stacked by radius.
The gray lines show the profile for individual voids, while the thick black
lines show the mean.
We chose to stack the voids in our catalog with radii in the range
$5 \leq r < 6 \, h^{-1}$Mpc, of which there are 511.
We note that the central density is about $0.2 \, \bar{\rho}$ in the dark
matter distribution, as a result of our choice of void finder.
The average profile then rises almost continuously to the void edge, though
individual voids show substructure within them
(also visible in Fig.~\ref{fig:slice}).
The slope of our profile contrasts with the profile of more evolved voids at
lower redshifts, which exhibit a `bucket' profile
\footnote{For example \citet{Cec06} propose
$\rho(r) / \bar \rho = A_0 + A_3 \left(r / R_V \right)^3$ while \citet{Ham15}
proposes a 4-parameter model with a similar shape.}.
We used the $z = 0$ simulation output to create a low-redshift SO void catalog
(with adjusted threshold and average target values) and found that these voids
do exhibit such a `bucket' profile.
The voids we are studying at $z = 2.5$, however, have not yet evolved to
such a state and are still in the process of evacuating.
The real-space flux profiles illustrate just how well the flux profile mirrors
the density profile. The center of the voids have $F \simeq 1$, which translates
to $\delta_F \simeq 0.25$ for our setting of $\langle F \rangle = 0.8$.
In both the density and flux value, the stack profile almost reaches the mean
value by $r = 10 \, h^{-1}$Mpc.

The middle panel of Fig.~\ref{fig:profile} shows the radial velocity profiles
of the voids.
The profiles are linearly increasing up to the void radius, where they peak
around $120 \, \mathrm{km \, s}^{-1}$ before turning over.
The radial velocity profiles also have a fairly small scatter -- at the average
radius of $5.4 \, h^{-1}$Mpc, the mean velocity is
$119.7 \, \mathrm{km\,s^{-1}}$ with a standard deviation of
$5.4 \, \mathrm{km\,s^{-1}}$ or about 5 per cent.
Within the context of linear theory
$\vec{v}(\vec{r}) \propto \vec{\nabla} \nabla^{-2} \delta(\vec{r})$.
If we assume a spherical mass distribution, this can be solved to yield
$v_r \propto \delta(<r) / r^2$, where $\delta(<r)$ is the ``overdensity enclosed
within $r$'' in analogy with the Newtonian gravitational acceleration due to
a spherical mass distribution. The radial velocity profile around
spherical, or averaged, voids then becomes \citep{Pee93, Ham15}\footnote{Note,
we have an additional factor of $a$ in this expression compared to
Equation~2.2 of \protect\citet{Ham15}, perhaps due to a difference in proper
vs.\ comoving quantities. We always use comoving scales and densities, and
peculiar velocities.}
\begin{equation}
  v(r) = -a f H \frac{1}{r^2}
    \int_0^r \left( \frac{\rho(x)}{\bar{\rho}}-1 \right) x^2 \, dx
\end{equation}
where $f \simeq \Omega_{\rm m}^{0.55}(z)$ is the growth factor, which at
$z = 2.5$ is close to 1.
This form was shown in fig.~1 of \citet{Ham15} to fit the velocity profile of
stacked voids in N-body simulations well at $z \simeq 0$.
The dashed red line in Fig.~\ref{fig:profile} shows this linear theory
approximation, which we see compares favorably to the profile measured in our
$z = 2.5$ voids (within 10 per cent over the range plotted).
It is somewhat surprising that the linear theory prediction matches our
simulated radial velocity profile result down to Mpc scales and for
$|\delta| = 0.8$.
The fact that this prediction also matched void radial velocity profiles at
$z = 0$, with voids from a different finder method is impressive \citep{Ham15}.

\begin{figure}
\begin{center}
\includegraphics[width=\columnwidth]{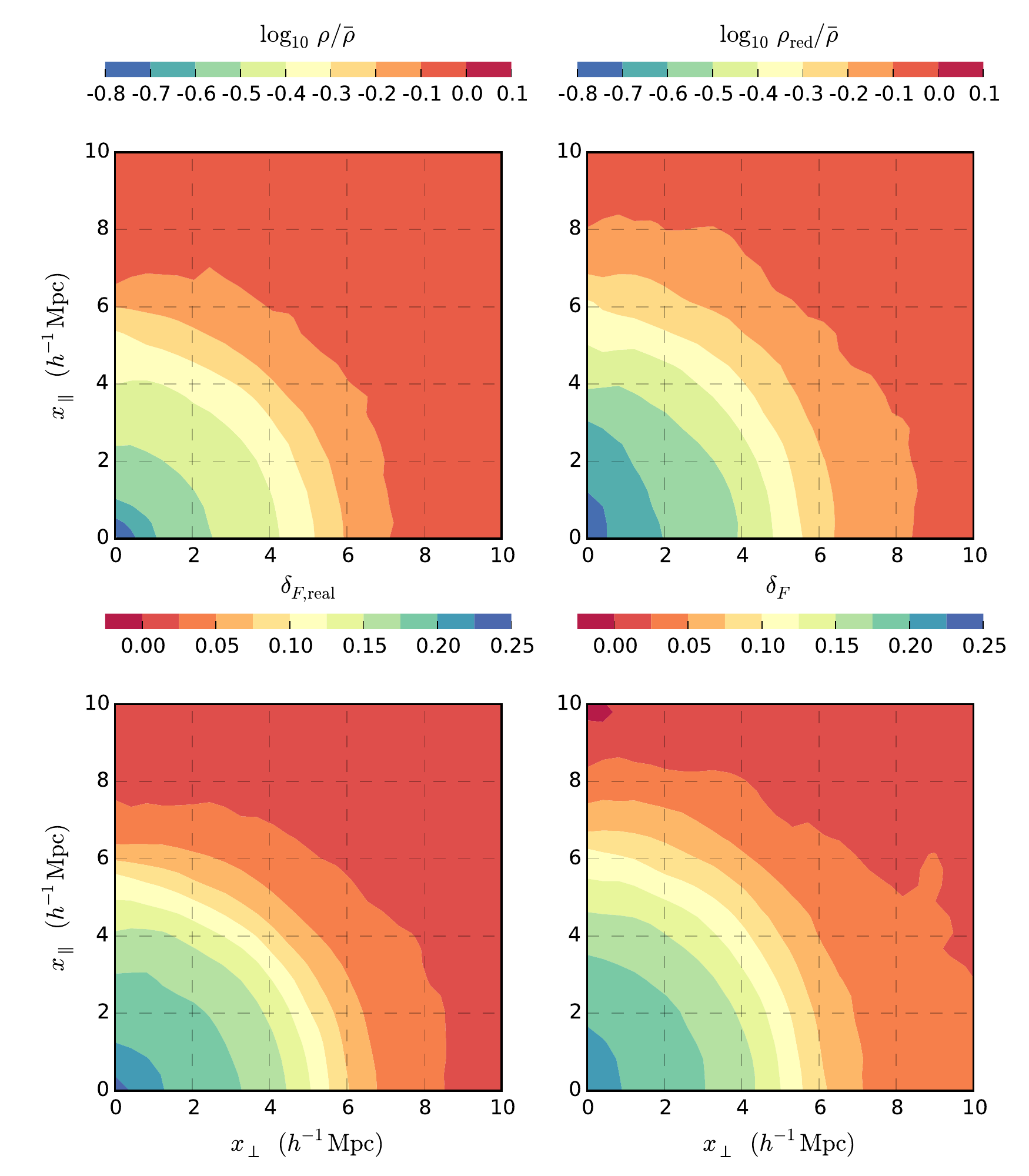}
\end{center}
\caption{
Contour plots of a stack of the voids with $5 \le r < 6 \, h^{-1}$Mpc,
showing the impact of redshift-space distortions.
The four panels show the density and flux in real and redshift space,
binned in the distances parallel ($x_\parallel$) and perpendicular ($x_\perp$)
to the line of sight.
In the redshift-space fields, the radial profiles are clearly extended in the
line-of-sight direction.
}
\label{fig:stack_rsd}
\end{figure}

Fig.~\ref{fig:stack_rsd} shows the two dimensional profiles (in mass and flux)
of stacks of voids with radii $5 \leq r < 6 \, h^{-1}$Mpc in both real and
redshift space.
Apart from some noise near the line-of-sight axes, the contours in
Fig.~\ref{fig:stack_rsd} are isotropic in the real-space panels but
show extended, anisotropic profiles in redshift-space.
This is an indication of the effect of peculiar velocities, which appear
visually to be larger in our case than at lower redshifts when voids
are traced by galaxies.
We note that the profiles are much better measured at small radii where there
is less scatter.
Beyond the radius of the stack $r \approx 5.5 \, h^{-1}$Mpc, the scatter in the
individual profiles increases significantly.
We believe this is the source of the extended orange contour in the bottom right
panel, for instance.

Since we expect the stacked voids to be isotropic in real space, by symmetry,
any observed anisotropy offers an opportunity to study such peculiar velocities.
This could be particularly interesting for constraining models with
modified gravity. For example, \citet{Cla13} find that, driven by the
outward-pointing fifth force, individual voids in chameleon models expand
faster and grow larger than in a $\Lambda$CDM universe.
Such effects would modify the profile of the stacked voids in a potentially
observable manner, allowing observations of voids in the Ly$\alpha$ forest
to test such models.
Based on the radial velocity profiles shown in Fig.~\ref{fig:profile} and the
measured standard deviation, one would need only about 20 voids with a radial
velocity measurement to reach one per cent standard error
(assuming Poisson errors).
With accurate enough radial velocity measurements from void anisotropies, it
should be possible to detect deviations at the 10 per cent level with
relatively small samples.
Conversely, the larger impact of redshift-space distortions in the
Ly$\alpha$ flux field means they must be modeled in order to make a
measurement of the Alcock-Paczynski effect \citep{Alc79} from stacked voids
in the flux field (see Section~\ref{subsec:ap}).

\section{Finding voids in flux}
\label{sec:flux}

\begin{table}
\begin{center}
\caption{Void catalogs}
\begin{tabular}{l r r r r r}
\hline
Field & SO Thresh. & SO Avg. & Count & Vol. Frac. \\
\hline
$\rho$ & $0.20 \, \bar \rho$ & $0.4 \, \bar \rho$ & 16,167 & 0.152 \\
$\rho_{\rm red}$ & $0.15 \, \bar \rho$ & $0.3 \, \bar \rho$ & 16,338 & 0.151 \\
$\delta_F$ & 0.224 & 0.167 & 16,296 & 0.150 \\
hires map & 0.224 & 0.167 & 16,586 & 0.203 \\
midres map & 0.224 & 0.167 & 8,724 & 0.181 \\
lores map & 0.224 & 0.167 & 5,565 & 0.153 \\
\hline
\label{tab:catalogs}
\end{tabular}
\end{center}
The spherical under/overdensity void catalogs used for comparison, found in our
$V \approx 1.7 \times 10^7 \, h^{-3} \mathrm{Mpc}^3$ simulation.
We use the original $\rho$ catalog as our `truth' and varied the SO parameters
for the $\rho_{\rm red}$ and $\delta_F$ catalogs to qualitatively match.
The hires map uses a $d = 2.5 \, h^{-1}$Mpc sightline spacing mock survey,
while the midres map comes from a $d = 4.0 \, h^{-1}$Mpc configuration,
and the lores map from a $d = 6.0 \, h^{-1}$Mpc configuration.
The noise and smoothing inherent in the tomographic reconstruction process
create larger differences in the catalog properties.
\end{table}

Underdense regions show up as high transmission regions in the 3D Ly$\alpha$
forest flux for $z = 2$ -- 3, as shown in Fig.~\ref{fig:slice} and
\ref{fig:profile}.
This is not necessarily the case at lower redshifts, since the characteristic
density probed by the forest increases with time \citep{Bec11, Luk15}.
At lower redshifts, it is difficult to see differences in transmission
passing through an underdense region vs.\ a moderately overdense region, since
it takes a significant overdensity to create an observable absorption feature.
Fortunately, there is a large overlap between the redshift range of the forest
accessible from the ground and the redshift range where mean density structures
scatter an observable fraction of the light.
Given this, finding voids in flux at $z = 2$ -- 3 is a matter of finding
coherent high-transmission regions.
In \cite{Sta14}, we outlined a simple method using tomographic flux maps to find
protoclusters (coherent low-transmission regions) at these redshifts.
We now adapt these methods to find coherent high-transmission regions,
corresponding to high-redshift voids.

The void catalogs used in this section are listed in Table~\ref{tab:catalogs},
including voids found in the redshift-space density, the flux, and three
tomographic flux maps.
In each case, we modified the SO threshold and average target parameters to
create a void catalog with roughly the same void count and radius distribution
as the real-space density catalog.
For reference, we use three of the tomographic flux maps created in
\citet{Sta14}.
These maps were constructed by mocking up a realistic survey covering the
simulation volume with signal-to-noise distributions similar to the pilot
observations of \citet{Lee14b}, and several other settings of the average
sightline spacing and minimum spectral signal-to-noise ratio.
The tomographic maps were then generated by running our reconstruction code on
the mock spectra.
The three maps we use are generated from mock surveys with average sightline
spacings of $d_\perp = 2.5$, 4, and $6 \, h^{-1}$Mpc.
The smallest sightline separation configuration is similar to the ongoing
CLAMATO survey, and the larger separation configurations are similar to what we
expect from large-area surveys on 8 -- $10\,$m telescopes like
the Subaru Prime Focus Spectrograph \citep[PFS; ][]{Tak14}
or the Maunakea Spectroscopic Explorer \citep{Sim14}.
We refer to these tomographic maps as the hires, midres, and lores flux maps.
We discuss the characteristics of the individual catalogs in the following
subsections and focus on our method to compare catalogs for now.
Comparing across void catalogs, we want to confirm that there are nearby pairs
of voids with similar radii.
We expect to find the same set of voids in density and flux,
as we have demonstrated how well-matched the fields are in previous sections.
This is mostly a matter of determining the best SO parameters for the flux.
The tomographic flux maps, however, are contaminated by and spectral noise in
individual mock spectra and shot noise due to sparse sampling of the field,
and this noise will certainly affect our capability of finding voids.

We use two metrics to compare the catalogs of voids found in different
fields.
The first metric is essentially the sum of the difference in the center
positions and radii, which we call the match error.
If we are comparing voids in catalog A with voids in catalog B,
for each A-B pair, we compute the match error
\[
  \epsilon = \frac{\sqrt{(r_A - r_B)^2 + |\vec{x}_A - \vec{x}_B|^2 / 3^2}}{r_A}
\]
where $r$ is the radius and $\vec{x}$ is the center position.
We chose this form of the error for several reasons.
First, this form of the error also allows for trading off differences in radii
and centers.
We want to consider the differences relative to the size of the void,
which will allow for larger center and radii differences for larger voids.
Note that this form of the error assumes the radius of void A is the reference.
Finally, we compare $1/3$ of the center difference to the radius difference
just due to the dimensionality (and empirically we found that the mean center
difference is about 3 times the mean radius difference).
Later in this section, we will show that a match error $\epsilon < 0.3$
qualifies as a good match for a void between two catalogs, and we will
use this cut to count which voids are `matched'.
The second metric we use is the total volume overlap between the voids in both
catalogs.
Clearly, this metric is less useful for telling if a catalog A void is
well-matched by a single catalog B void.
However, it is a useful measure of how well-matched the catalogs are
overall and does not depend on a specific form of the error nor a specific value
to cut at.
It is also useful in cases where noise in the tomographic maps artificially
combines or splits voids -- although the centers and radii might not match
across catalogs, there will still be a sizable volume overlap.
We can use these metrics to get a sense of void completeness and purity of
each of the flux catalogs with respect to the density void catalog.
The number of `matched' density voids compared to the total number of density
voids (the match fraction) is a measure of completeness.
We also measure completeness by comparing the overlapping volume between two
catalogs to the total volume in density voids (the overlap fraction).
The purity of the flux catalogs can be measured by matching in the other
direction (the fraction of matched flux voids) and by comparing the overlap
to the total volume in flux voids.

Using these metrics, we first found that redshift-space distortions can create
large differences in the centers and, to a lesser extent, the radii of the
voids.
In order to more easily compare voids found in density and flux,
we created a void catalog using the redshift-space density.
Voids found in redshift-space density matched those found in real-space density
best (in terms of detecting the same voids with similar radii)
when we used a threshold of $0.15 \, \bar \rho$ and an average target of
$0.3 \, \bar \rho$. These densities are lower than the real-space values
since outflows from voids drive densities lower.

Before applying these metrics to the void catalogs derived from the various flux
maps, we also compared the redshift-space density void catalog to random void
catalogs, mainly to get a sense of the worst-case performance.
We created ten catalogs of 16,338 voids (the same number as the redshift-space
density catalog), with centers uniformly distributed in the simulation
domain, and with radii randomly drawn from the same distribution as that in
the redshift-space density catalog.
We compared each random void catalog against the density catalog, computing
the fraction of density voids with a match error $\epsilon < 0.3$ and the
fraction of the total volume overlap to the total volume in density voids.
Overall, 2.7 per cent of the density voids were matched by voids in the random
catalogs on average.
It is reassuring to see that a small fraction of the density voids are matched
by random voids which tells us that our cut of $\epsilon < 0.3$ is stringent
enough.
We also noticed that for the largest voids ($r \geq 8 \, h^{-1}$Mpc),
the average match fraction drops to 1.3 per cent.
This is due to the fact that both the density and random catalogs contain just
a few very large voids and it is even less likely that they will overlap enough
to meet the match error cut.
The average volume overlap fraction between density voids and voids in the
random catalogs was 15 per cent, and did not change with the radius considered.
This is not surprising since the voids cover roughly 15 per cent
of the total volume, so random points will overlap about that often.

\begin{table*}
\begin{center}
\caption{Catalog comparison for $r \geq 6 \, h^{-1}$Mpc voids}
\begin{tabular}{l c c c c c c}
\hline
           & Density        & Flux           & Hires map      & Midres map     & Lores map      & Random \\
\hline
Density    & --             & 0.994 / 0.937  & 0.660 / 0.631  & 0.478 / 0.568  & 0.269 / 0.517  & 0.0194 / 0.152 \\
Flux       & 0.988 / 0.933  & --             & 0.683 / 0.637  & 0.471 / 0.569  & 0.292 / 0.514  & 0.0188 / 0.153 \\
Hires map  & 0.581 / 0.567  & 0.576 / 0.567  & --             & 0.356 / 0.484  & 0.238 / 0.430  & 0.0261 / 0.153 \\
Midres map & 0.284 / 0.409  & 0.282 / 0.408  & 0.258 / 0.425  & --             & 0.171 / 0.344  & 0.0243 / 0.152 \\
Lores map  & 0.186 / 0.349  & 0.184 / 0.347  & 0.191 / 0.377  & 0.192 / 0.359  & --             & 0.0204 / 0.153 \\
Random     & 0.0193 / 0.151 & 0.0191 / 0.150 & 0.0368 / 0.204 & 0.0480 / 0.182 & 0.0398 / 0.149 & --             \\
\hline
\label{tab:cat_comp}
\end{tabular}
\end{center}
\begin{flushleft}
The catalog void match fraction and volume overlap fractions
(separated by a slash in each cell), for $r \geq 6 \, h^{-1}$Mpc voids.
For each row, we compute the fraction of voids with a match error
$\epsilon < 0.3$ and the fraction of the total volume overlapped by voids
in the catalog of that column.
\end{flushleft}
\end{table*}

\subsection{Ideal flux}

We ran our SO void finder on the `true' flux grid ($256^3$) using a threshold of
$\delta_F \geq 0.224$ and an average target of $\delta_F = 0.167$.
We experimented with several values of the threshold and average target fluxes
and found that these values resulted in a number of voids and radius
distribution similar to the catalog of voids found in density.
The mapping from flux to density evolves quickly with redshift, so these SO
parameters would have to be adjusted for other redshifts and UV background
prescriptions.

The flux void catalog matches the redshift-space density catalog very well.
For all voids ($r \geq 2 \, h^{-1}$Mpc), 84 per cent of the density voids and
the flux voids are matched.
The volume overlap fraction is also very high, at 86 per cent of the
density void volume and also 86 per cent of the flux void volume.
For larger voids ($r \geq 6 \, h^{-1}$Mpc), the catalogs are even better
matched.
In this case, 99.4 per cent of the density voids and 98.8 per cent of the
flux voids are matched, and the volume overlap fraction is 93.7 per cent
of the density voids and 93.3 per cent of the flux voids.
For reference, the density catalog contains 335 of these large voids,
while the flux catalog contains 325.

\subsection{Tomographic flux maps}

We constructed the map void catalogs by running the SO void finder on the maps
with the same SO parameters we used for the ideal flux field.
We tried several other SO parameter settings on the maps, but found that the
default value catalogs performed best in comparison to the density and flux
catalogs.
Small changes to the SO parameters resulted in slightly better performance, but
changes larger than about $\Delta \delta_F = 0.01$, resulted in similar or worse
performance, so we did not bother optimizing these parameter choices further.
Unfortunately, our simple void-finding method does not consider noise in the
map which can contaminate the set of thresholded points and the spherical
averages.
The noise in the map acts to scatter points below or above the SO threshold,
creating false negatives and positives respectively.
In the same way, the noise can affect the spherical averages used in the SO
finder, resulting in inaccurate radii.
However, this should be less of an issue for coherent structures spanning
several map resolution scales, which is apparent in our results for small vs.\
large voids.
The effects of the noise are apparent in the radius distribution of the
map catalogs.
In the hires map catalog, the number of very small voids
($r < 2.5 \, h^{-1}$Mpc) is 3,902, about half of the number found in the
density catalog (6,157).
This is likely due to shot noise where sightlines did not sample these smaller
structures well enough.
The number of medium voids
($3 \, h^{-1} \mathrm{Mpc} \leq r < 6 \, h^{-1}$Mpc) is about double that in the
density catalog, and the number of large voids ($r \geq 7 \, h^{-1}$Mpc) about
the same (147 vs.\ 121).
This explains why there is a similar total number of voids in the density and
hires map catalogs, but more total volume in the map catalog
(see Table~\ref{tab:catalogs}).
The radius distributions of the midres and lores map catalogs are more distorted
by the noise.
The midres map catalog contains about a half the number of small voids
($r < 5 \, h^{-1}$Mpc) compared to the density catalog and the lores map
catalog contain about a quarter.
There are approximately double the number of large voids
($r \geq 7 \, h^{-1}$Mpc) in both map catalogs compared to the density catalog.

\begin{figure}
\begin{center}
\includegraphics[width=\columnwidth]{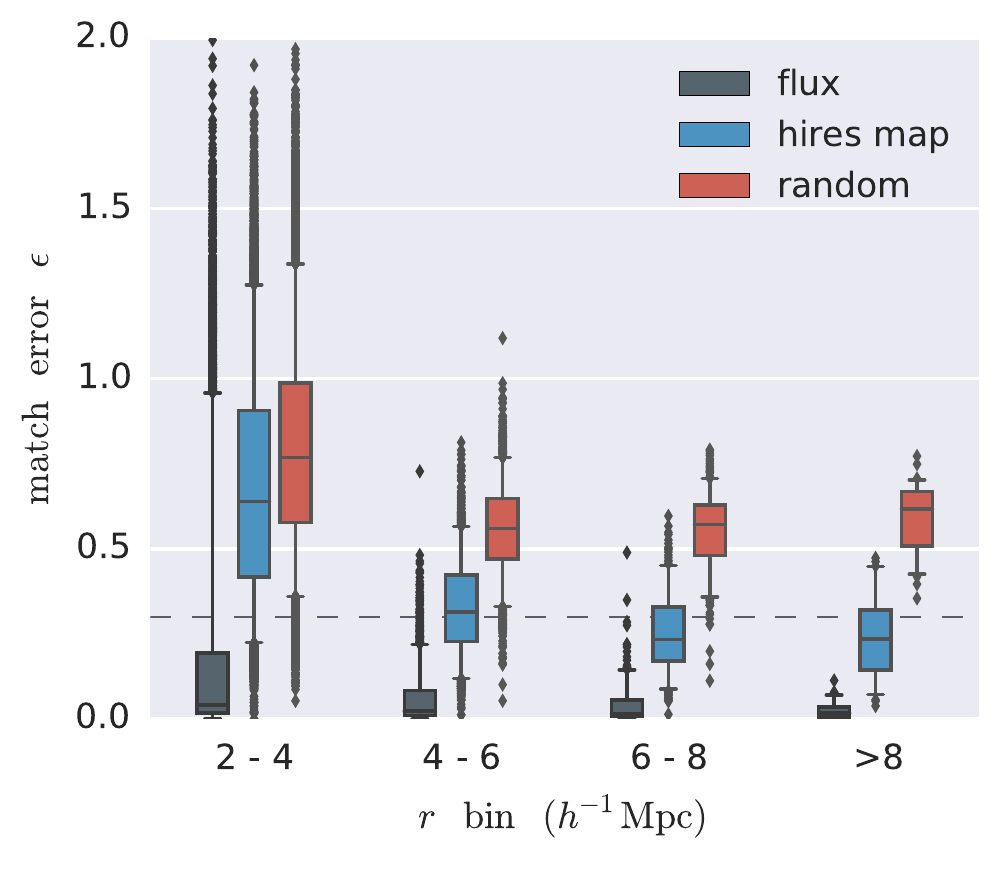}
\end{center}
\caption{
A box and whisker plot of the distributions of match errors vs.\ radius.
The match error is computed between the redshift-space density voids and the
voids in the flux catalog (gray), the hires map catalog (blue), and
in one of the random catalogs (red).
The line in the middle of the box is the median, the box extends from the 25th
to the 75th percentile, the whiskers extend down to the 5th and up to the 95th
percentiles, and points outside of this range are plotted individually.
The match error threshold of $\epsilon = 0.3$ is marked with a dashed line.
The flux catalog matches the density catalog exceptionally well.
The hires map catalog is essentially random for small voids,
but performs much better for large voids.
The random catalog match error is fairly flat across radius bins.
}
\label{fig:match_errors}
\end{figure}

In Fig.~\ref{fig:match_errors}, we plot distributions of the match errors
between the voids in the redshift-space density catalog and voids in the flux,
hires map, and one of the random catalogs vs.\ radius.
The line in the middle of the box shows the median, the box extends from the
25th to the 75th percentiles, the whiskers extend from the 5th to the 95th
percentile, and samples outside this are plotted individually.
The match error cut value of $\epsilon = 0.3$ is marked with a dashed line.
The match errors against the random catalog tend to fall around
$\epsilon = 0.6$, and there are few points under our cut of $\epsilon = 0.3$,
again showing that this is a safe choice.
It is also reassuring that the random errors are relatively flat over radius
bins because we defined the match error relative to the original void radius.
The gray distributions show just how well-matched the flux and density catalogs
are and that the flux match errors overlap very little with the random errors.
Overall, the hires map catalog misses a significant fraction of the small voids
in the density catalog, but performs well for larger voids.
For all voids ($r \geq 2 \, h^{-1}$Mpc), the hires map catalog matches only
16.3 per cent of the density voids and the volume overlap fraction is
49.4 per cent.
This can also be seen in the smallest radius bin in Fig.~\ref{fig:match_errors},
where the hires map is just a bit lower than the random distribution.
However, considering larger voids ($r \geq 6 \, h^{-1}$Mpc), the hires map
catalog performs much better matching against 66 per cent of the density voids
and overlapping with 63 per cent of the volume.
In Fig.~\ref{fig:match_errors}, there is a clear trend that the hires map
match errors decrease with radius, separating from the random distribution.
The match fractions in the other direction (purity of the hires map voids) are
similar at 17.8 per cent for all voids and 58.0 per cent for large voids.
The lower match fractions for large voids in this case is driven by the hires
map catalog having more large voids.

The midres map void catalog performs worse for all voids, but still matches a
considerable fraction of the density voids.
Overall, the midres map catalog matches only 6.2 per cent of the density voids,
although it still overlaps with 40 per cent of the density void volume.
If we consider larger voids ($r \geq 6 \, h^{-1}$Mpc), the midres map catalog
matches 48 per cent of the density voids, and matches 60 per cent of even
larger voids ($r \geq 8 \, h^{-1}$Mpc).
The lores map catalog performs worse than this, but is still useful for finding
large voids.
The lores map catalog matches only 2.5 per cent of the density voids
-- consistent with the random catalogs -- although it overlaps with 30 per cent
of the volume.
This indicates that the void finder is still able to find regions containing
voids from the map, but does not recover an accurate center or radius.
If we consider some of the largest voids ($r \geq 8 \, h^{-1}$Mpc), the
match fraction increases to 48.9 per cent and the volume overlap fraction is
61.9 per cent, again confirming that the larger the void,
the better the maps perform.

Our results are also summarized in Table~\ref{tab:cat_comp}.
In this table, we give the match and overlap fractions between the
redshift-space density, flux, hires map, midres map, lores map,
and random catalogs for the voids with $r > 6 \, h^{-1}$Mpc.
We note again that the first five catalogs are single catalogs while the random
results are averages over the ten random catalog realizations.
The trends between the catalogs are the same as described above:
the random catalogs match 2 -- 3 per cent and overlap about 15 per cent,
and the correspondence between the density and flux catalogs is very high.
Comparing the density (or flux) catalogs to the map catalogs, the match fraction
drops to 60 -- 70 per cent for the hires case, to 40 -- 50 per cent for the
midres case, and down to 20 -- 30 per cent for the lores case.
However, the volume overlap fraction remains relatively high for all of the maps
indicating that the poor matching is more the fault of our simple void finding
method than the maps truly missing the voids.

\begin{figure*}
\begin{center}
\includegraphics[width=6in]{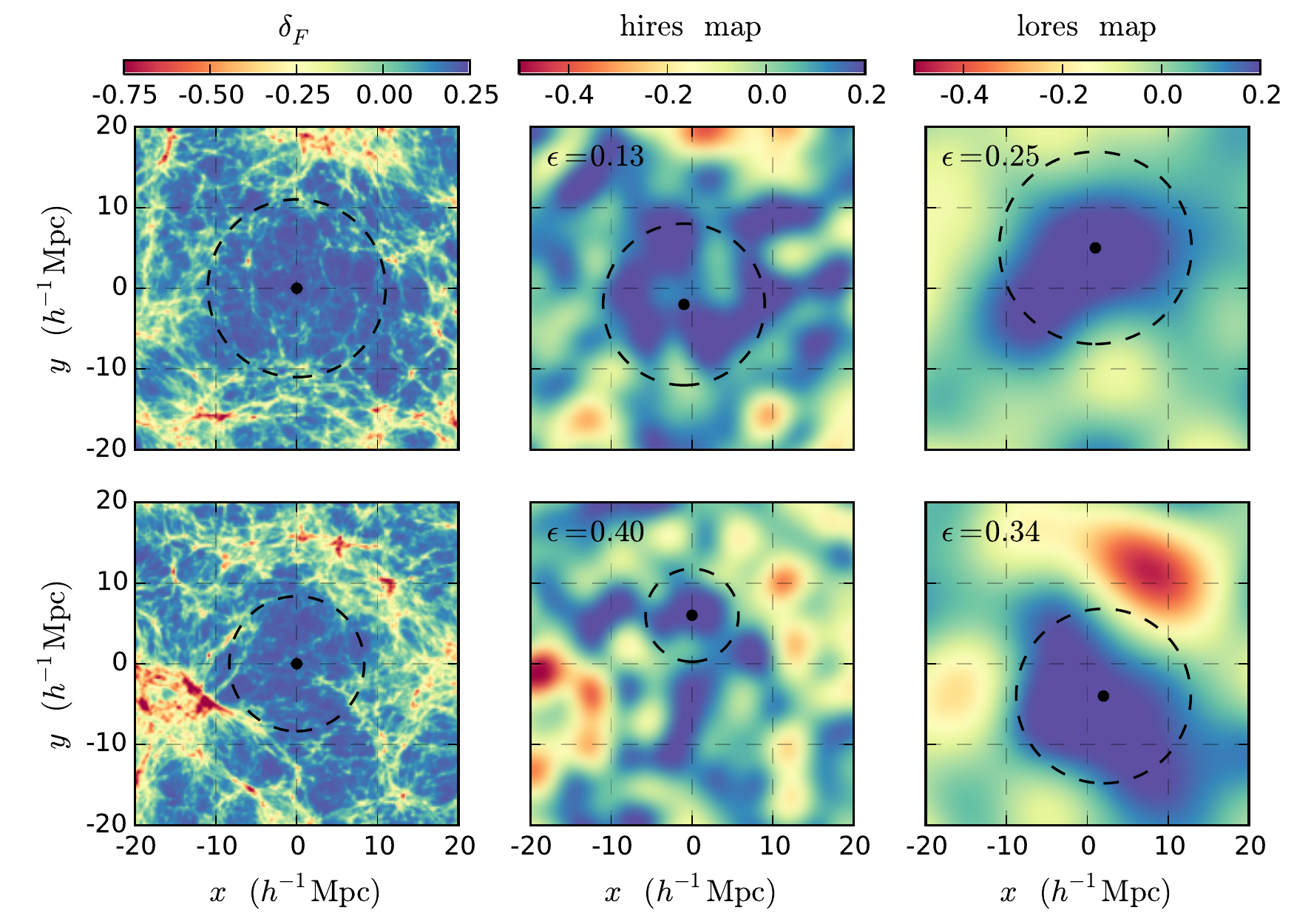}
\end{center}
\caption{
Slices of two large voids showing the flux, hires map, and lores map (from left
to right).
A void that is well-matched in the map void catalogs is shown on top,
and a poor match case is shown on the bottom.
The top void has a radius of $10.9 \, h^{-1}$Mpc and the bottom void has a
radius of $8.3 \, h^{-1}$Mpc.
Just as in Fig.~\ref{fig:slice}, the slice is $40 \, h^{-1}$Mpc across and
$6 \, h^{-1}$Mpc into the page, although in this case, the redshift direction is
into the page.
The black dot and circle in each panel show the void as found in each catalog.
The flux panels show the original void (very close to the one found in density),
while the map panels show the map voids with the lowest match error.
We also annotate the match error of the map voids in the top-left corner of the
images.
}
\label{fig:map_slices}
\end{figure*}

Overall, the maps perform decently matching voids with radii larger than the
map resolution, but it is surprising that the maps still do not perform better
for the largest voids.
We visually inspected many of the large voids to see why the flux map void
catalogs sometimes miss these large voids.
We show two example slices of large voids in the flux, hires map, and lores
map in Fig.~\ref{fig:map_slices}.
The top row shows a successful void match in both maps, while bottom row shows
a failure case.
In the flux panels, we overplot the original void with a black dot and circle.
In the map panels, we overplot the best match void and annotate the match error.
In the top panels, the matching hires map void is a bit smaller and offset just
a bit to the bottom left.
In the lores map, a noise feature around $(x, y) = (5, -5) \, h^{-1}$Mpc
pushes the matching void center up farther, but with a similar radius,
resulting in a sufficiently small match error.
In the bottom hires map panel, there is significant noise around the center of
the original void which pushes the center of the void up and restricts the
growth of the void radius to a much smaller size.
In the bottom lores map panel, the filamentary structure around
$(x, y) = (5, -5) \, h^{-1}$Mpc is missing, which allows the void radius to
grow much larger and results in a poor match.
Interestingly, in the bottom row, the overall structure of the lores map matches
the structure of the ideal flux better than the hires map by eye.
However, we have not considered estimate of the noise in the map.
Using the tomographic reconstruction method outlined in \citet{Sta14}, it is
possible to compute the covariance of the map or to run Monte Carlo error
estimates.
Considering the amount of noise apparent in Fig.~\ref{fig:map_slices}, much
could be gained by incorporating a noise estimate into a void finding procedure.
We believe future work can make significant gains in void finding performance
by considering the structure of voids beyond simple spheres and taking the map
noise into account.

\section{Discussion}
\label{sec:discussion}

\subsection{Survey prospects}

For the cosmology of our simulation, the comoving radial distance to $z = 2.5$
is $4050 \, h^{-1}$Mpc, thus one degree subtends $70 \, h^{-1}$Mpc.
Assuming a $250 \, h^{-1}$Mpc depth (e.g.\ $2.2 < z < 2.5$), each square degree
of survey area translates into a volume of
$1.2 \times 10^6 \, h^{-3} \mathrm{Mpc}^{-3}$.
Given the number densities in Fig.~\ref{fig:radius_dist}, we see that
surveys like CLAMATO with $V \simeq 10^6 \, h^{-3} \mathrm{Mpc}^3$
would encompass about $150$ voids larger than $5 \, h^{-1}$Mpc in radius.
If we assume a conservative void finding efficiency of 60 per cent, our
simple method would recover $\sim 90$ voids.
This would be the first detection of a significant population of high-redshift
voids.
Of course, this is a lower bound on the efficiency of identifying voids with
a map of this resolution due to our conservative choice of what constitutes a
match, and that there is still room for improvement in the method.
Using the Subaru Prime Focus Spectrograph \citep[PFS; ][]{Tak14},
it is possible to double the target density, covering a larger redshift range
at the cost of sightline density.
In \citet{Lee14a}, we discussed piggybacking on the planned galaxy evolution
survey described in \citet{Tak14}.
Such a survey would provide a map of roughly $16 \, {\rm deg}^2$ or
$8 \times 10^{4} \, h^{-2} \mathrm{Mpc}^2$ area and $700 \, h^{-1}$Mpc depth
($2.3 < z < 3.2$) for a total volume of
$6 \times 10^{7} \, h^{-3} \mathrm{Mpc}^3$, although at a coarser resolution of
about $5 \, h^{-1}$Mpc.
This much larger volume would encompass $\sim 3000$ voids with
$r \geq 5 \, h^{-1}$Mpc, and would detect voids with an efficiency better
than 30 per cent, providing a sample of around $\sim 1000$ voids.
With an extended program on PFS of 100 nights, it is possible to construct a
tomographic map covering $\sim 200 \, \mathrm{deg}^2$ with the same redshift
coverage and resolution, providing a tenfold increase in volume, and therefore,
the number of voids ($\sim 10^4$).

For comparison, similar volumes have been explored to find voids in
low-redshift galaxy positions, although for somewhat larger voids.
\citet{Pan12} searched for $r > 10 \, h^{-1}$Mpc voids in the
Sloan Digital Sky Survey Data Release 7 main galaxy sample (out to $z = 0.1$),
corresponding to a volume $V \approx 10^7 \, h^{-1} \mathrm{Mpc}^3$, finding
$\sim 1000$ voids.
\citet{Sut12} also found a similar number of voids in the SDSS DR7 main galaxy
sample (out to $z = 0.2$) and the luminous red galaxy sample
(out to $z = 0.44$).
In total, the galaxy samples were split into 6 samples covering volumes
from $10^6$ to almost $10^9 \, h^{-1}$Mpc (see their table~2).
However, the larger volume samples were covered by brighter, more massive
galaxies, with larger separations.
By $z = 0.1$, the average galaxy separation in DR7 is already larger than
$5 \, h^{-1}$Mpc, making it difficult to find statistically significant
small galaxy voids.
\citet{Sut14a} provided an update to this analysis using the Baryon Oscillation
Spectroscopic Survey Data Release 9 CMASS sample, split into 6 samples,
each covering about $5 \times 10^8 \, h^{-3} \mathrm{Mpc}^3$.
This work found $\sim 1500$ voids with large radii ($> 20 \, h^{-1}$Mpc).

\subsection{High-Redshift Void Cosmology}
\label{subsec:ap}

\begin{figure}
\begin{center}
\includegraphics[width=\columnwidth]{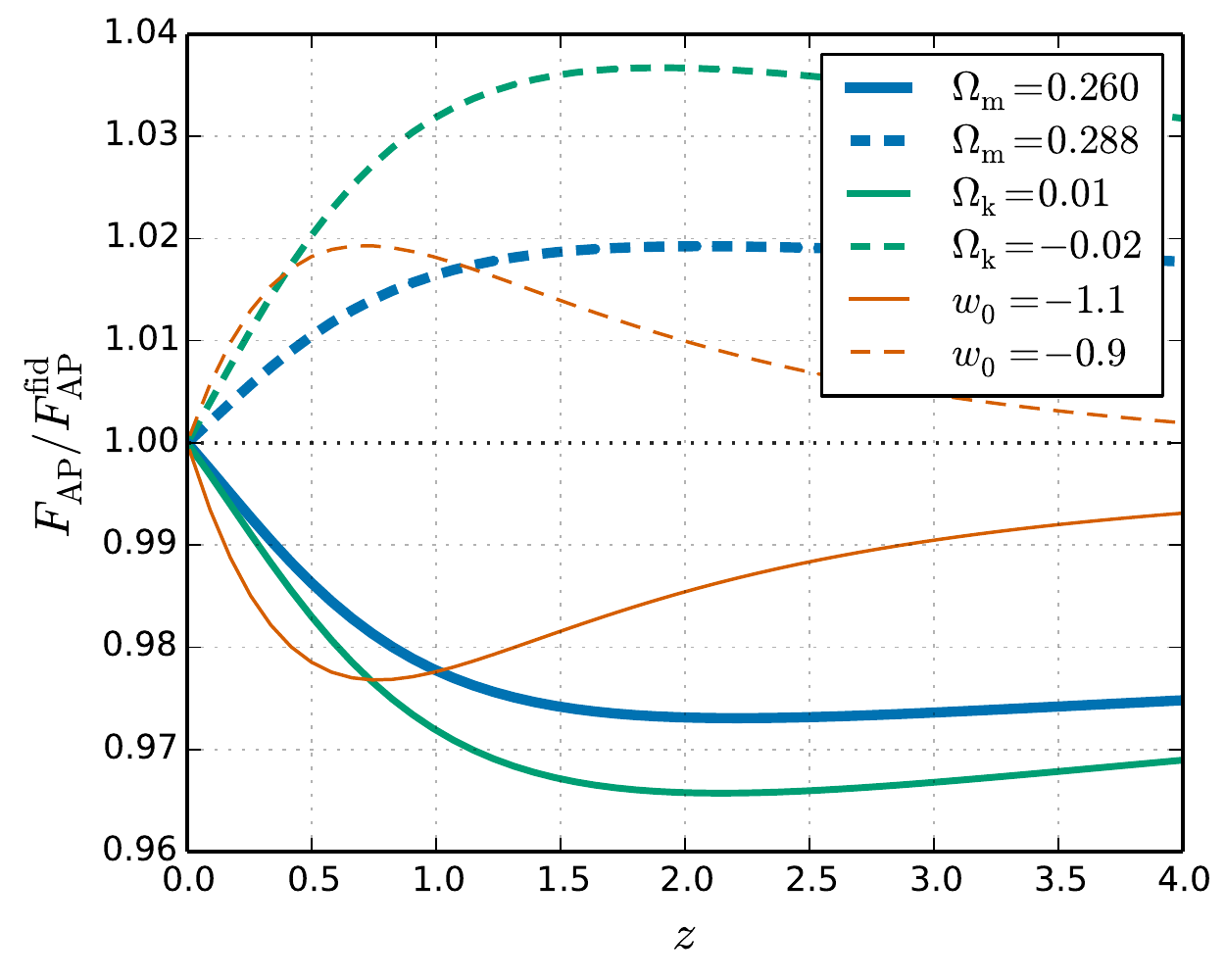}
\end{center}
\caption{
$F_{\rm AP}$ as a function of redshift for different cosmological models,
divided by the prediction from our fiducial model.
We show 5 per cent variations in $\Omega_{\rm m}$, 1 or 2 per cent variations in
$\Omega_{\rm k}$ per cent, and 10 per cent variations in $w_0$.
For all models, the value of the Hubble parameter $h$ has been adjusted to keep
the angular scale of the CMB oscillations $\theta_{\rm ls} = d_{\rm ls} / r_s$
fixed.
State-of-the-art measurements of $F_{\rm AP}$ have 5 per cent uncertainties,
measured at $z = 0.5$ up to $z = 2.4$
\protect\citep{Bla11, Aub14, Beu14, Sam14}.
}
\label{fig:f_ap}
\end{figure}

Much of the recent discussion of voids as cosmological probes has focused on
them as a means to measuring the Alcock-Paczynski (AP) parameter,
\begin{equation}
  F_{\rm AP} = \frac{1 + z}{c} D_{\rm A} H,
\end{equation}
where $D_{\rm A}$ is the angular diameter distance and the Hubble parameter,
$H$, encodes distortions in the line-of-sight (LOS) direction.
Note that this measurement measures $H(z)$ directly, rather than an integral as
measured by e.g.\ Type Ia supernovae.
In Fig.~\ref{fig:f_ap}, we show how specific variations in cosmological
parameters affect the AP parameter, giving a qualitative idea of how accurate
these measurements must be.
Specifically, we show changes in the AP parameter with 5 per cent variations in
$\Omega_{\rm m}$, 1 or 2 per cent variations in $\Omega_{\rm k}$ per cent,
and 10 per cent variations in $w_0$.

\citet{Lav12} have proposed measuring the AP parameter using the anisotropy of
stacked voids, in the context of voids identified in galaxy surveys.
Even though voids should be spherically symmetric when averaged, each void will
have a certain random asymmetry that will add noise to the global stack.
\citet{Lav12} showed that the uncertainty due to this intrinsic scatter when
averaging $N$ voids can be approximated by (their eq.~35):
\begin{equation}
 \frac{\sigma_{AP}}{F_{\rm AP}} \approx \frac{1}{\sqrt{N}} ~.
 \label{eq:sigmaAP}
\end{equation}
In terms of cosmological parameters, Fig.~\ref{fig:f_ap} shows that it requires
substantial changes in cosmological parameter values, by today's standards,
to produce one or two per cent changes in $F_{\rm AP}$.
Therefore, in order to be competitive with other cosmological probes,
the stack should be done using several thousand voids.

In order to accurately estimate the uncertainty when stacking $N$ voids
identified in the flux field, we would have to study the intrinsic scatter in
the asymmetry of the void flux profiles, as a function of redshift and void
size.
We would also have to take into account the effect of potential systematics
like errors in centering and measuring radii, as well as different sources of
contamination in the Ly$\alpha$ flux.
But assuming that we would also need ten thousand voids to have a one per cent
measurement, we can use the discussion above to estimate that we could
achieve this uncertainty with a CLAMATO-like survey over 100 square degrees, or
with a PFS-like survey over 200 square degrees.

Moreover, $F_{\rm AP}$ can also be robustly constrained from anisotropic
measurements of the Baryon Acoustic Oscillation (BAO) scale
\citep{Eis98, Seo03}.
BAO measurements typically report ratios of separations with respect to a
fiducial model along the line of sight
($\alpha_\parallel \pm \sigma_\parallel$) and transverse
($\alpha_\perp \pm \sigma_\perp$) directions, as well as their
correlation coefficient ($r$).
One can translate these values into a ratio of $F_{\rm AP}$ with respect to
$F_{\rm AP}$ in the fiducial model:
\begin{equation}
  f_{\rm AP} = \frac{F_{\rm AP}}{F_{\rm AP}^{\rm fid}}
    = \frac{\alpha_\perp}{\alpha_\parallel}
\end{equation}
with an uncertainty given by:
\begin{equation}
  \frac{\sigma_f^2}{f_{\rm AP}^2} =
    \frac{\sigma_\parallel^2}{\alpha_\parallel^2}
    + \frac{\sigma_\perp^2}{\alpha_\perp^2}
    - 2 \frac{r \sigma_\parallel \sigma_\perp}{\alpha_\parallel \alpha_\perp} ~.
\end{equation}
For instance, recent BAO measurements from the BOSS collaboration
\citep{Aub14, Sam14} can be translated into $\sim 5$ per cent measurements of
$f_{\rm AP}$ both at $z = 0.57$ (from the galaxy survey) and at $z=2.4$
(from the Ly$\alpha$ survey), raising the bar for measurements from voids.

\section{Conclusions}
\label{sec:conclusions}

In this paper, we characterized the signal of cosmological voids in the
high-redshift matter density field and demonstrated how we can use
Ly$\alpha$ forest tomographic maps to find high-redshift voids.
We used a simple spherical over/underdensity approach to identifying voids in
a large cosmological simulation (with a box size of $256 \, h^{-1}$ Mpc or a
volume of $1.7 \times 10^7 \, h^{-3} \mathrm{Mpc}^3$) at $z = 2.5$, resulting in
a catalog of $\sim 16,000$ voids with radii of 2 -- $12 \, h^{-1}$Mpc.
We also tested finding voids with a watershed approach and found that the
resulting catalog was similar to that produced by the spherical overdensity
method, but with more complex geometries that changed the void centroid
non-trivially.
For simplicity, we used the spherical overdensity void finding method
throughout.
This makes our results somewhat conservative, i.e.\ it is likely that more
sophisticated void-finding methods will have improved performance.

Overall, the signature of high-redshift voids in flux is similar to what has
been found for low-redshift voids in density.
The radial density profile of voids is low ($\rho / \bar \rho = 0.2$ -- 0.4)
and rises more steeply closer to the radius of the void.
One difference we noticed is that the high-redshift voids are typically less
evacuated than their low-redshift analogues, giving them a steeper inner profile
and less pronounced rise at the edge.
The shape of the density profile is clearly mirrored in flux with high
transmission inside the radius ($\delta_F = 0.25$ -- 0.15), and dropping down
to the mean flux beyond the radius.
Interestingly, the radial velocity profiles show very little scatter and the
mean radial velocity profile matches up to the linear theory prediction very
well.
This could be a promising testbed for any (modified gravity) theory predicting
differences in void outflow velocities.

Using our void finding method, we identified voids in an ideal flux field
and in three tomographic flux maps generated from mock surveys with
spatial samplings of $d_\perp = 2.5$, 4, and $6 \, h^{-1}$
(hires, midres, and lores maps).
We compared the flux void catalogs to the density void catalogs by considering
how well `matched' pairs of voids are in terms of their centers and radii.
We found excellent agreement between the density and ideal flux void catalogs,
where 99 per cent of the large voids ($r > 6 \, h^{-1}$Mpc) are well-matched.
The noise in the tomographic maps clearly impacts the efficiency of finding
voids, reducing the fraction of well-matched large voids down to 66, 48, and 27
per cent in the hires, midres, and lores maps, respectively.
However, when we inspected individual cases of poorly matched voids, we found
that many of these are due to noise in the maps artificially breaking up or
merging high-transmission regions.
It is clear that a more sophisticated void finder, especially one that models
a noise component, would perform much better on the tomographic maps.
Implementing such a method is beyond the scope of the current work.

Using these matching results, we can provide a conservative forecast for the
number of voids that can be found in dense Ly$\alpha$ surveys.
Our hires map has a signal-to-noise ratio distribution and sightline spacing
similar to the ongoing CLAMATO survey.
With a sky coverage of one deg$^2$, the CLAMATO data would produce a
tomographic map covering $V \approx 10^6 \, h^{-1} \mathrm{Mpc}^3$,
and our proposed void-finding method would identify about 100 voids with
$r > 5 \, h^{-1}$Mpc in such a volume.
With a $16 \, \mathrm{deg}^2$ survey on the PFS, we would identify about 1000
voids with $r > 5 \, h^{-1}$Mpc, although at a degraded purity.
A 100-night dedicated Ly$\alpha$ forest survey across $200 \, \mathrm{deg}^2$
on the PFS would increase this number by a further order of magnitude to
$>10^4$ voids.

These populations of high-redshift voids could be useful for many purposes,
including tests of modified gravity, as an AP test and for studying
high-redshift void galaxies.
Previous works have considered voids as a clean environment for studying
galaxy evolution, where galaxies are very isolated and their evolution is not
complicated by environmental effects \citep[e.g.\ see][sec.~5]{Wey11a}.
However, existing studies of void galaxies are concentrated at low redshift,
where such objects are much easier to find \citep{Wey11b}.
At low redshifts, the evidence points to the different properties of void
galaxies being caused by their low stellar mass, independent of other
influence from their void environment \citep{Hoy05, Kre11, Tin08}.
It would be very interesting to see whether similar behavior is seen at
higher redshifts, where we expect the processes of galaxy formation could be
different.
Current galaxy redshift surveys can probe only down to $L \sim L_\star$
in galaxy luminosity at these redshifts, and we would naively expect
high-redshift voids identified through Ly$\alpha$ forest tomography to
also be void of such bright galaxies.
However, the James Webb Space Telescope and its NIRSPEC
spectrograph\footnote{\url{http://www.stsci.edu/jwst/instruments/nirspec}}
will have the ability to target $L \sim 0.3 L_\star$ galaxies within voids
identified through CLAMATO and PFS.

With dense Ly$\alpha$ forest surveys covering larger volumes, such as a
dedicated program on the PFS covering $200\,\mathrm{deg}^2$,
it is possible to identify a population of $10^4$ voids.
Such a large number of voids would naively translate to a one per cent AP
measurement, although this is just a statistical estimate and it is
possible that there would be larger systematic errors in such a measurement.

As described in App.~\ref{app:data}, we are providing public access to the data
used in this project including gridded simulation quantities,
the tomographic flux maps, a grid of hires flux skewers, FoF halo catalogs,
void catalogs, and the protocluster catalog from \citep{Sta14}.

We thank Kathryn Kreckel for useful discussions.
The simulation, mock surveys, and reconstructions discussed in this work were
performed on the Edison Cray XC30 system at the National Energy Research
Scientific Computing Center, a DOE Office of Science User Facility supported by
the Office of Science of the U.S. Department of Energy under Contract No.
DE-AC02-05CH11231.
This research has made use of NASA's Astrophysics Data System
and of the astro-ph preprint archive at arXiv.org.

\bibliographystyle{mn2e}
\bibliography{ms}

\appendix

\section{Comparison of SO and watershed voids}
\label{app:sheds}

\begin{figure}
\begin{center}
\includegraphics[width=\columnwidth]{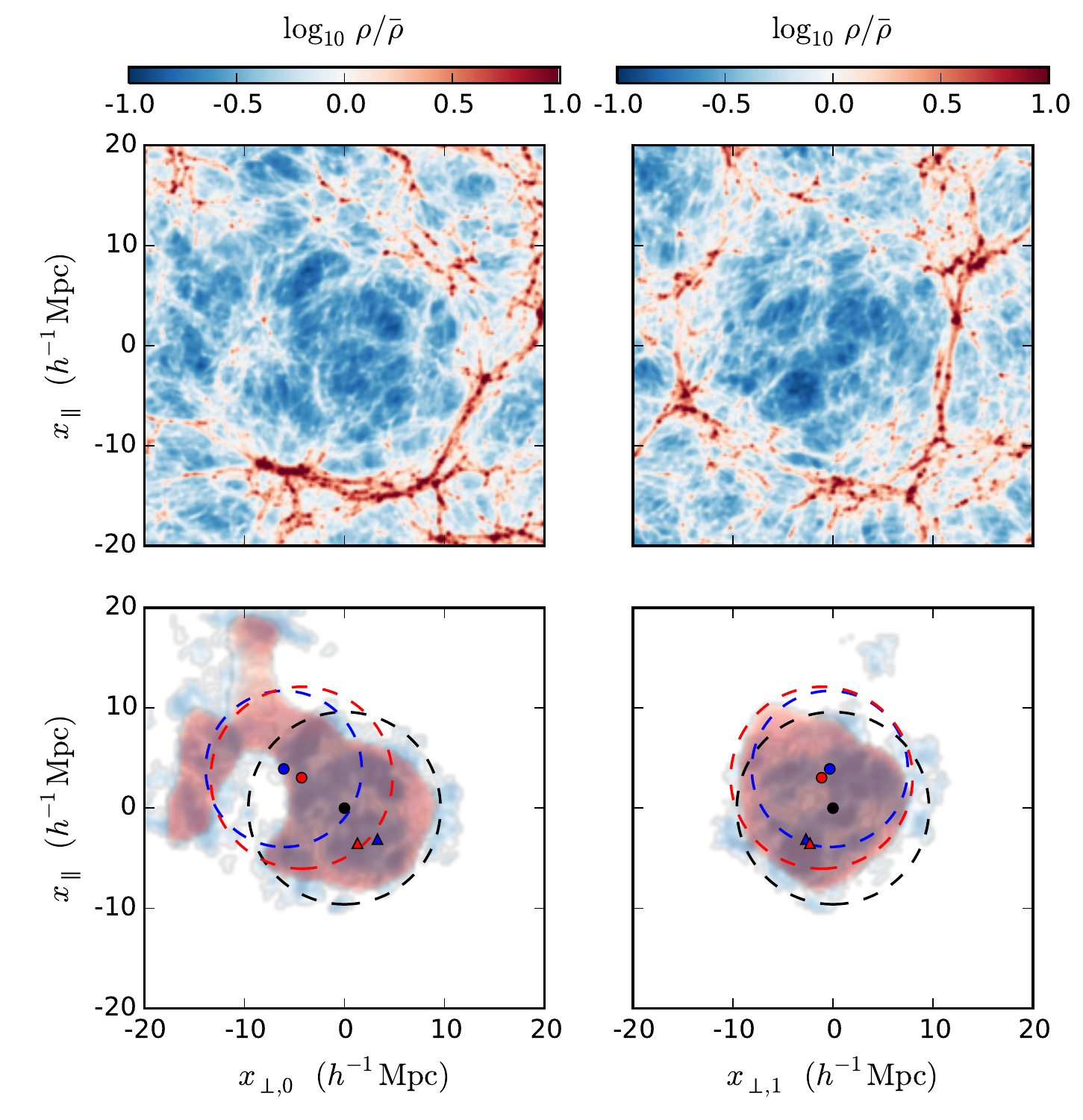}
\end{center}
\caption{
A density slice centered on a large void with a radius of $9.6 \, h^{-1}$Mpc.
Just as in Fig.~\ref{fig:slice}, the slice is $40 \, h^{-1}$Mpc across and
$6 \, h^{-1}$Mpc into the page.
The panels on the left show the view along the simulation $y$ axis,
while the panels on the right show the view along the simultion $x$ axis.
(Top) The matter density.
(Bottom) The corresponding voids from the SO catalog (black),
the watershed catalog (blue), and the smoothed density watershed catalog (red).
The dots and dashed line circles show the center of the voids and the SO radius
or watershed effective radius.
The triangles show the minimum value (core) points of the watersheds.
The blue and red colorscales show the projection of the points in the watershed
(the darker the color, the more points into the page).
See the text for more details.
}
\label{fig:watershed}
\end{figure}

In Subsection~\ref{subsec:void_finding}, we briefly compared our SO void catalog
to a set of voids found via a watershed method with the same threshold setting.
We concluded that it would be more straightforward to use the voids found via
the SO method, mainly due to the complex shapes of the watershed.
In this Appendix, we show an example void found in both catalogs to illustrate
this point.
In Fig.~\ref{fig:watershed}, we show a slice centered on a large void.
The left and right panels show the same void structure from two angles
(the xz-plane and yz-plane).
The top panels show the density field in this region, while the bottom panels
show the void shape in the different catalogs.
The black dot and circle are the center and radius of the void found with the
SO method.
The blue dot and circle show the value-weighted centroid and effective radius
of the void found with the watershed method,
and the blue triangle shows the `core' point
(the minimum value point within the shed).
We also show the points in the void watershed in the bottom panels with the blue
colorscale, where the color scales with the number of points in the projection.
In order to damp out some of the complex structure of the original watershed
void, we also tried running the watershed finder on a $2 \, h^{-1}$Mpc smoothed
density field (with an adjusted threshold of $0.45 \, \bar \rho$).
The watershed void found in the smoothed field is shown in red.

The SO and watershed voids have reasonably similar shapes as seen in the
yz-plane (right panel).
The extent of the watershed points (blue region) overlaps almost entirely with
the SO circle (black), besides the small wayward blue blob at
$(y, z) = (5, 15) \, h^{-1}$Mpc.
However, seen in the xz-plane, the voids have very different shapes indeed.
The slightly overdense region at $(x ,z) = (-10, 0) \, h^{-1}$Mpc limits the
growth of the SO void, but the watershed region reaches around this structure
to the underdense region on the other side.
This extension from the main underdense region is also seen in the smoothed
version of the watershed void.
The SO radius is $9.6 \, h^{-1}$Mpc and the watershed effective radius is
$7.8 \, h^{-1}$Mpc, and $9.1 \, h^{-1}$Mpc in the smoothed version.
Although the radii are all fairly similar, it's amazing to see just how
different the extents and centers differ.
The SO center is $7.2 \, h^{-1}$Mpc away from the watershed centroid and
$5.3 \, h^{-1}$Mpc away from the watershed core.
At the same time, the watershed core and centroid are separated by a whopping
$11.9 \, h^{-1}$Mpc, far more than the effective radius.
Overall, the watershed void spans 39, 26, and $32 \, h^{-1}$Mpc in the
x, y, and z directions respectively, meaning the small finger-like voids
extending from the central underdensity are very long.

\section{Public data products}
\label{app:data}

The data used in this project are available at
\url{http://tinyurl.com/lya-tomography-sim-data}.
We hope that making this data publicly available will reduce the barrier to
future work on Ly$\alpha$ forest tomography and high-redshift voids.
The data release includes gridded simulation quantities, the tomographic flux
maps, a grid of hires flux skewers, FoF halo catalogs, void catalogs, and
the protocluster catalog from \citep{Sta14}.
Due to space limitations, we downsampled the gridded quantities from the
full $2560^3$ grid to a $640^3$ grid.
Although this process erases some small-scale structure, the resolution is still
more than enough for our purposes.
The gridded quantities include the $z = 2.5$ density, redshift-space density,
flux, real-space flux, and peculiar velocities and the $z = 0$ density and
peculiar velocities.
We also include example Python and C++ sources for reading the files.

\label{lastpage}

\end{document}